\documentclass[11pt]{article} 

\usepackage{graphicx}
\usepackage{psfig,picinpar,floatflt,amssymb,epsf}
\usepackage[english]{babel}
\usepackage{latexsym}
\usepackage{amsmath,amsfonts,amssymb,amsthm}
\def\bseq{\begin{subequation}}  % = 1a 1b
\def\eseq{\end{subequation}}
\def\bsea{\begin{subeqnarray}}  % = 1.1a 1.1b
\def\esea{\end{subeqnarray}}
\newcommand{\bbox}{\lower.2ex\hbox{$\Box$}}

\newcommand{\beq}{\begin{equation}}
\newcommand{\eeq}{\end{equation}}
\newcommand{\bea}{\begin{eqnarray}}
\newcommand{\eea}{\end{eqnarray}}
\newcommand{\ena}{\end{eqnarray}}

\newcommand {\non}{\nonumber}

\renewcommand{\a}{\alpha}
\renewcommand{\b}{\beta}
\renewcommand{\c}{\chi}
\newcommand{\cb}{\bar{\chi}}
\renewcommand{\d}{\delta}
\newcommand{\pa}{\partial}

\newcommand{\g}{\gamma}
\newcommand{\G}{\Gamma}

\newcommand{\z}{\zeta}

\newcommand{\mb}{\overline{m}}

\newcommand{\F}{\Phi}

\newcommand{\Fib}{\bar{\Phi}}

\newcommand{\s}{\sigma}
\renewcommand{\S}{\Sigma}

\newcommand{\te}{\theta}

\newcommand{\sba}{\bar{\sigma}}
\newcommand{\Sb}{\bar{\Sigma}}

\newcommand{\Tr}{{\rm Tr}}
\renewcommand{\(}{\left(}
\renewcommand{\)}{\right)}
\renewcommand{\[}{\left[}
\renewcommand{\]}{\right]}
\newcommand{\ggl}{\left\{}
\newcommand{\ggr}{\right\}}

\newcommand{\Db}{\bar{D}}

\newcommand{\Fb}{\bar{F}}

\newcommand{\psib}{\bar{\psi}}

\newcommand{\ad}{{\dot{\alpha}}}
\newcommand{\bd}{{\dot{\beta}}}

\newcommand{\Dc}{\nabla}
\newcommand{\Dcb}{\bar{\nabla}}

\begin{document}

\begin{titlepage}
{\hbox to\hsize{April 2004 \hfill}}
\begin{center}
\vglue .06in
\vskip 40pt
%\vskip .9in
{\Large\bf Quantization of $N=1$ chiral/nonminimal (CNM) scalar multiplets and 
supersymmetric Yang--Mills theories} 
\\[.7in]
{\large\bf Gabriele Tartaglino Mazzucchelli\footnote{gabriele.tartaglino@mib.infn.it}}
\\[.5in]
{\it Dipartimento di Fisica, Universit\`a degli studi di
Milano-Bicocca\\ 
and INFN, Sezione di Milano, piazza delle Scienze 3, I-20126 Milano, Italy}
\\[.7in]

{\bf ABSTRACT}\\[.3in]
\end{center}
We give the superfield quantization of chiral/nonminimal (CNM) scalar 
multiplets defined by pairs of $N=1$ chiral and complex linear scalar
superfields kinematically coupled. In the pure massive case we develop the 
covariant quantization when CNM multiplets are coupled to background gauge 
superfields. Furthermore, we study some properties of $N=1$ supersymmetric 
Yang--Mills theories constructed using CNM scalar matter superfields. In 
particular, we compute the one--loop contribution to the effective action for 
the matter superfields, we study the analogue of the Konishi anomaly and 
discuss some properties of the glueball superpotential.

\vskip 30pt
${~~~}$ \newline
\\[.02in]  

\end{titlepage}

\section{Introduction}

Generally, the $N=1$ irreducible scalar chiral multiplet, defined by the 
superfield $\F$ satisfying $\Db_\ad\F=0$, is used to construct
four--dimensional supersymmetric models having matter contents given by $N=1$ 
scalar multiplets. However, other less studied nonminimal off--shell 
representations of the $N=1$ scalar multiplet can be found in the literature
\cite{SUPERSPACE,NM,CNM}. 
While all these representations describe the dynamics of spin $(0,1/2)$ 
physical fields \cite{CNM,CNMmassless}, they differ from the chiral scalar 
multiplet in the auxiliary fields content.
In particular, among these nonminimal multiplets the complex linear 
superfield presents a number of interesting properties.

The complex linear superfield, defined by the kinematic constraint 
$\bar{D}^2\Sigma=0$, appears in various contexts in the superspace description
of supersymmetric field theories. It is present, for example, as a conformal 
compensator in different formulations of supergravity
\cite{CompensatoriConformi}, and naturally appears in the context of N=2 
off-shell supersymmetric sigma models \cite{CNMN=2}. 
Furthermore, in contrast with other nonminimal 
scalar multiplet representations, it can be easily coupled to Yang--Mills 
fields \cite{CNM,NM.S.Y.M.}.
The massless complex linear superfield also possesses interesting 
properties of duality with the massless chiral superfield 
\cite{SUPERSPACE,CNM,Sigma+PropLinear,NM.S.Y.M.}.\\
In the literature, the Dirac spinor is usually embedded in $N=1$ SUSY theories
using a pair of chiral multiplets. 
However, an alternative realization of a Dirac spinor in $N=1$ SUSY theories
makes use of one chiral and one complex linear superfield 
\cite{CNM,CNMmassless}, and this kind of construction 
differs from the pure chiral case in some interesting aspects.
For example, it was observed in \cite{CNMmassless} 
that the formulation of the Dirac spinors using chiral and complex linear 
multiplets gives gauge group transformation properties for the Dirac spinors 
which are holomorphic vector--like. Moreover, it was also seen 
that the chiral/nonminimal Dirac spinor could provide a 
solution to the propagation of auxiliary fields in 
$N=1$ supersymmetric extensions of the low energy effective QCD actions, 
a problem which arises naturally in the formulation in terms of chiral 
multiplets. This kind of models also provide a new way of realizing
parity violation \cite{CNMmassless}. 
Therefore, the complex linear superfield could be a relevant tool in the 
formulation of phenomenological $N=1$ supersymmetric models \cite{CNMmassless}.

To define a consistent supersymmetric mass term it is possible 
to build a SUSY model for a chiral superfield $(\F)$ and a complex linear 
superfield $(\S)$ coupled through a modification
of the complex linear superfield kinematic definition: 
$\Db^2\S=Q(\F)=m\F+\F\widetilde{P}(\F)$ \cite{CNM}\footnote{A different way to
give mass to the complex linear superfield without introducing chirals is 
discussed  in \cite{BuchbinderKuzenko}.}. This coupling 
gives the same mass $m$ to the two multiplets and produces a nontrivial 
interaction. Therefore, the complex linear multiplet can acquire a mass 
$m$ ``in tandem'' with a chiral multiplet through the previous definition 
\cite{CNM}. Moreover, this kinematic constraint 
does not break the natural Dirac spinor construction of the two 
superfields \cite{CNM}.  
Following \cite{CNMmassless}, we call these models chiral/nonminimal (CNM)
models.

In order to study some properties of CNM models in this letter we quantize 
the CNM superfields in superspace with generic $Q(\F)$. To do this we  
generalize to the present case the known quantization techniques for the 
massless chiral \cite{SUPERSPACE} and complex linear 
\cite{Q.F.T.S.NM,Sigma+PropLinear} cases using unconstrained superfields
solving the kinematic constraints. When the mass parameter 
is strictly $m\ne 0$ and the CNM multiplets are coupled to background gauge
fields, we build the covariant quantization.

Once given the quantization, we then construct $N=1$ Super--Yang--Mills 
theories with CNM matter superfields. In particular, for the CNM multiplet we 
consider the simplest case $\Db^2\S=m\F$. Taking into account the
propagation of both the matter and SYM vector superfields, we  
compute the one--loop contribution to the effective action for the matter
superfields $\S$ and $\F$.\\
Using the covariant formalism we then derive the analogue of 
the Konishi anomaly \cite{Konishi} in CNM theories finding, as argued in 
\cite{CNMmassless}, that the CNM theory is anomaly free. 

The letter is organized as follow: In section $2$ we give a brief description
of CNM models. In section $3$ we develop the quantization of CNM theories,
while in section $4$ we calculate the one--loop contribution
to the effective action for the matter fields $\F$ and $\S$ in SYM theories
with CNM matter superfields. 
In section $5$ we first discuss the covariant quantization of CNM multiplet
coupled to background gauge fields and then apply the covariant formalism
to the study of the Konishi anomaly. In section $6$ we discuss some properties
of the glueball superpotential in CNM SYM theories. In section $7$ we present 
some final remarks concerning the dynamics of CNM models.

For the conventions adopted see reference \cite{SUPERSPACE}.

\section{Chiral/nonminimal (CNM) scalar models}
\label{section:Q.CNM}

In this section we introduce models built through $N=1$ chiral and nonminimal
scalar multiplets \cite{NM,SUPERSPACE} and we consider the possibility of 
coupling these multiplets in accordance with \cite{CNM}.\\ 
In particular, using the $N=1$ superspace formalism \cite{SUPERSPACE}, we 
consider a chiral superfield $\F$ satisfying $\Db_\ad\F=0$ and, for the 
nonminimal scalar multiplet, take a complex linear superfield $\S$ 
satisfying $\Db^2\S=0$ \cite{NM,CNM,SUPERSPACE}. Separately the two multiplets
have the kinetic actions
\bea
S_{C}=\int d^4x d^4\te~ \Fib\F&,&S_{NM}=-\int d^4x d^4\te~ \Sb\S~~~.
\eea
In components these have the form \cite{NM,SUPERSPACE}
\bea
S_{C}&=&\int d^4x [\bar{A}\Box A-\bar{\psi}^\ad i\pa_{\a\ad}\psi^\a+\Fb F~]
~~~,\\
S_{NM}&=&\int d^4x [\bar{B}\Box B -\bar{\zeta}^\ad i \pa_{\a \ad} \zeta^\a
-\bar{H} H +\beta^\a \rho_\a +\bar{\rho}^\ad \bar{\b}_\ad - \bar p^{\a \ad}
p_{\a \ad}~]~~~.
\eea
From these expressions it is possible to see that both actions describe the 
free dynamics of two $N=1$ scalar multiplets with physical fields given by 
$(A,\psi_\a)$ for the chiral scalar multiplet, and by $(B,\z_\a)$ for the 
complex linear multiplet. Clearly, the two multiplets have different auxiliary
field contents.

It is possible to introduce interaction terms between these multiplets
described by 
\beq
S_{int}=\int d^4x d^4\te~ K(\F,\Fib,\S,\Sb)+\ggl\int d^4x d^2\te~ W(\F)+ 
{\rm h.c.}\ggr~,
\label{Sint}
\eeq
where $K$ is the K\"ahler potential (at least cubic) 
and $W$ is a holomorphic function of the chiral superfield $\F$ only.\\
There is also the possibility to introduce a mass term and
a nontrivial interaction between the two multiplets $\F$ and $\S$ by 
modifying the kinematic constraints for the superfields as 
\cite{CNM}
\bea
\Db_\ad\F=0&,&\Db^2\S=Q(\F)~~~,\non\\
D_\a\Fib=0&,&D^2\Sb=\bar{Q}(\Fib)~~~,
\label{defkinCNM}
\eea
where $Q(\F)$ is a holomorphic function of the chiral superfield $\F$.
In this letter we consider $Q(\F)$ of the form 
\bea
Q(\F)\equiv
%\F\widetilde{Q}(\F)=\F\[m+\sum_{k=1}^n a_k\F^k\] \equiv 
\F\[m+\widetilde{P}(\F)\]~,
\label{QCNM}
\eea
where $\widetilde{P}(\F)$ is polynomial in $\F$ at least linear.\\
The simplest action for the CNM models \cite{CNM} is then
\bea
S_{CNM}&=&\int d^4xd^4\te \[~\Fib\F-\Sb\S~\]~~~.
\label{SCNM}
\eea
Due to the constraints (\ref{defkinCNM}), in components (\ref{SCNM}) takes the 
nontrivial form \cite{CNM}
\bea
S_{CNM}&=&\int d^4x \Big[\bar{B}\Box B+\bar{A}\Box A 
- \bar{\zeta}^\ad i \pa_{\a \ad} \zeta^\a-
\bar{\psi}^\ad i\pa_{\a \ad}\psi^\a+\bar{F}F
-\bar{H} H +\beta^\a \rho_\a +\bar{\rho}^\ad \bar{\b}_\ad+ \non\\
&&- \bar p^{\a \ad}p_{\a \ad}
-\bar{Q}(\bar{A})Q(A)
-\big\{[B(\bar{Q}'(\bar{A})\bar{F}+\frac{1}{2}
\bar{Q}''(\bar{A})\bar{\psi}^\ad\bar{\psi}_\ad)+Q'(A)\zeta^\a\psi_\a]+
{\rm h.c.}\big\}\Big]~,~
\label{SCNMcomp}
\eea
with $Q'(A)={\pa Q\over \pa A}$ , $Q''(A)={\pa^2 Q\over \pa A^2}$. 
In the particular case $Q(\F)=m\F$ we have
\bea
S^0_{CNM}&=&\int d^4x [\bar{B}\Box B+\bar{A}\Box A 
- \bar{\zeta}^\ad i \pa_{\a \ad} \zeta^\a-
\bar{\psi}^\ad i\pa_{\a \ad}\psi^\a+\bar{F}F
-\bar{H} H +\beta^\a \rho_\a +\bar{\rho}^\ad \bar{\b}_\ad+ \non\\
&&~~~~~~- \bar p^{\a \ad}p_{\a \ad}
-|m|^2\bar{A}A-\{(\mb B\bar{F}+m\zeta^\a\psi_\a)+
{\rm h.c.}\}]~,
\label{SquadCNM}
\eea
which, after integration of the auxiliary fields, gives
\bea
S^0_{CNM}&=&\int d^4x [\bar{B}(\Box-|m|^2) B+\bar{A}(\Box-|m|^2) A
+\frac{1}{2}\overline{\Psi}_{CNM}(i\g^\mu\pa_\mu-m)\Psi_{CNM}]~.
\label{SdiracCNM}
\eea
Here $\Psi_{CNM}$ is the Dirac spinor
\bea
\Psi_{CNM}\equiv \(\begin{array}{c} \psi_\a \\\bar{\z}^\ad \end{array}\)=
\(\begin{array}{c} D_\a\F| \\\Db^\ad\S| \end{array}\)~.
\label{diracCNM}
\eea
From (\ref{SdiracCNM}) it follows that the chiral/nonminimal 
multiplets acquire the same mass. Thus they naturally define Dirac spinors in 
$N=1$ supersymmetric theories \cite{CNM}. This interesting property
was considered in \cite{CNMmassless} in order to study 
$N=1$ supersymmetric extensions of QCD effective actions, and it is one of the 
main reasons to consider Super--Yang--Mills theories with CNM 
matter multiplets.
In fact, it is not difficult to extend the previous construction to CNM scalar 
multiplets minimally coupled to gauge multiplets; we need only to replace the
superspace covariant derivatives with derivatives 
$\Dc_A\equiv (\Dc_\a,\Dcb_\ad,\Dc_{\a\ad})$ covariant under supersymmetry and 
gauge transformations \cite{SUPERSPACE,CNM}. 

To conclude this section we consider the possibility of also introducing in 
(\ref{SquadCNM}) the quadratic chiral mass term 
$-{m'\over 2}\int d^4x d^2\te~ \F^2 +{\rm h.c.}=-\int d^4x
\{m'(AF+\psi^2)+{\rm h.c.}\}$, typical of pure chiral theories.
Integrating out the auxiliary fields the resulting action is
\bea
&&\int d^4x[\bar{B}\Box B+\bar{A}\Box A 
- \bar{\zeta}^\ad i \pa_{\a \ad} \zeta^\a-\bar{\psi}^\ad i\pa_{\a \ad}\psi^\a
-(|m|^2+|m'|^2)\bar{A}A-|m|^2\bar{B}B
-m'\mb AB+\non\\&&~~~~~~-\mb' m\bar{A}\bar{B}
-m\z^\a\psi_\a-m'\psi^2-\mb\bar{\z}^\ad\psib_\ad-\mb'\psib^2]~.
\label{Scnm+c}
\eea
The bosonic mass matrix is not diagonal in this case. It is possible to
diagonalize the bosonic field equations by a unitary constant bosonic field 
redefinition
\bea
&\(\begin{array}{c}B\\\bar{A}\end{array}\)=
\(\begin{matrix}{|m|\over \sqrt{|m|^2+|\widetilde{m}_1|^2}} &&
{|m|\over \sqrt{|m|^2+|\widetilde{m}_2|^2}}\\\\
{\mb(|\widetilde{m}_1|^2-|m|^2)\over
\mb'|m|\sqrt{|m|^2+|\widetilde{m}_1|^2}}&&
{\mb(|\widetilde{m}_2|^2-|m|^2)\over
\mb'|m|\sqrt{|m|^2+|\widetilde{m}_2|^2}}
\end{matrix}\)
\(\begin{array}{c}\widetilde{B}\\\widetilde{\bar{A}}\end{array}\)~~~,
\non\\\non\\
&|\widetilde{m}_{1,2}|^2=|m|^2+\frac{|m'|}{2}\(|m'|\pm \sqrt{|m'|^2+4|m|^2}\)
~~~,
\eea
$|\widetilde{m}_{1,2}|^2$ being the eigenvalues of the bosonic mass matrix.
Therefore (\ref{Scnm+c}) can be written as
\bea
\int d^4x \Big{[}\widetilde{\bar{B}}(\Box-|m_1|^2)\widetilde{B}+
\widetilde{\bar{A}}(\Box-|m_2|^2)\widetilde{A}+
\frac{1}{2}\overline{\Psi}_{CNM}(i\g^\mu\pa_\mu-m)\Psi_{CNM}+\non\\
+\big{[}-\frac{m'}{2}\overline{\Psi}_{CNM}^c\big{(}{1+\g_5 \over 2}\big{)}
\Psi_{CNM}+{\rm h.c.}\big{]}\Big{]}~~~,~~~~~~~~~~~~~~
\label{Scnm+ctilde}
\eea
where $\Psi_{CNM}^c=\(\begin{array}{c}\z_\a\\\bar{\psi}^\ad\end{array}\)
={\cal C}\overline{\Psi}_{CNM}^t$ is the charge conjugate spinor of 
$\Psi_{CNM}$ (\ref{diracCNM}) being 
${\cal C}$
%$=\(\begin{matrix}-C_{\a\b}&0\\0&C^{\ad\bd}\end{matrix}\)$ 
the charge conjugation matrix \cite{SUPERSPACE,BuchbinderKuzenko}.
From (\ref{Scnm+ctilde}) we observe that the resulting fermion mass matrix 
breaks the Dirac spinor construction.
Therefore, in this letter we assume $m'\equiv0$ and focus only on actions 
where the chiral interaction potential $\[\int d^6z W(\F)+{\rm h.c.}\]$ is at
least cubic in the chiral superfields.

\section{Quantization of CNM models with generic coupling}
\label{section:QCNM1}

We consider the action (\ref{SCNM}) with a pair of
superfields $\F$ and $\S$ satisfying the kinematic constraints
(\ref{defkinCNM}). We also assume in the action an 
interaction term that is local and at least cubic in the fields $\F$, $\Fib$, 
$\S$ and $\Sb$ as given in (\ref{Sint}). 

Our goal is to develop the superspace quantization of the model 
for a generic potential $Q(\F)$ as in (\ref{QCNM}) generalizing the 
quantization procedure of 
the massless complex linear superfield \cite{Q.F.T.S.NM,Sigma+PropLinear}.\\
Since, unlike the chiral scalar superfield \cite{SUPERSPACE}, an explicit 
formulation of the functional differentiation and integration for the 
superfields $\S$ and $\Sb$ is not known, we solve the kinematic constraints 
which define $\F$ and $\S$ through two unconstrained superfields $\c$ and 
$\s_\a$
\bea
\F\equiv \Db^2\c~~~&,&~~~\S=\Db^\ad\sba_\ad+m\c+\c\widetilde{P}(\Db^2\c)~,
\label{risvincoloCNM1}\\
\Fib\equiv D^2\cb~~~&,&~~~\Sb=D^\a\s_\a+\mb\cb
+\cb\overline{\widetilde{P}}(D^2\cb)~.
\label{risvincoloCNM2}
\eea
In terms of $\c$, $\cb$, $\s_\a$ and $\sba_\ad$ superfields action 
(\ref{SCNM}) reads
\beq
S_{CNM}=\int d^4xd^4\te\[(D^2\cb)(\Db^2\c)-\(D^\a\s_\a+\mb\cb+
\cb\overline{\widetilde{P}}(D^2\cb)\)\(\Db^\ad\sba_\ad+m\c+
\c\widetilde{P}(\Db^2\c)\)\]~.
\label{Siniz}
\eeq
Once the kinematic constraints have been solved, varying the superfields $\c$, 
$\cb$, $\s_\a$ and $\sba_\ad$ in the action $S_{CNM}+S_{int}$, we obtain the
classical equations of motion
\bea
&&\Db_\ad\[-\Sb+{\pa K(\F,\Fib,\S,\Sb)\over \pa \S}\]=0~~~,~~~
D_\a\[-\S+{\pa K(\F,\Fib,\S,\Sb)\over \pa \Sb}\]=0~~~,~~~~~~~~~~~~
\non\\\non\\
&&\[-\Sb+{\pa K(\F,\Fib,\S,\Sb)\over \pa \S}\]{\pa Q(\F)\over \pa\F}+
\Db^2\[\Fib+{\pa K(\F,\Fib,\S,\Sb)\over \pa \F}\]+{\pa W(\F)\over\pa\F}=
0~~~,~~~~~~~~~~~\non\\\non\\
&&\[-\S+{\pa K(\F,\Fib,\S,\Sb)\over \pa \Sb}\]{\pa \bar{Q}(\Fib)\over\pa\Fib}+
D^2\[\F+{\pa K(\F,\Fib,\S,\Sb)\over \pa \Fib}\]+
{\pa \overline{W}(\Fib)\over\pa\Fib}=0~~~.~~~~~~~~~~~~\label{eqCNM}
\eea
From these equations it follows that on--shell 
$\[-\Sb+{\pa K(\F,\Fib,\S,\Sb)\over \pa \S}\]$ defines a class of composite
chiral operators. In particular, if ${\pa K(\F,\Fib,\S,\Sb)\over 
\pa \S}\equiv0$, the $\Sb$ ($\S$) superfield becomes on--shell chiral 
(antichiral).

The quadratic part of action (\ref{Siniz}) is 
\bea
S^0_{CNM}
&=&\int d^8z(~\cb~,~\s_\a~)\(\begin{matrix}(D^2\Db^2-|m|^2)&-\mb\Db^\ad\\
-mD^\a&-D^\a\Db^\ad \end{matrix}\)\(\begin{array}{c}\c\\
\sba_\ad\end{array}\)~.
\label{S0CNM}
\eea
The kinetic operator is not invertible, since $\c$, $\cb$, $\s_\a$ and 
$\sba_\ad$ in (\ref{risvincoloCNM1}, \ref{risvincoloCNM2}) are 
defined up to two sets of gauge transformations which leave $\F$, $\Fib$, 
$\S$, $\Sb$ invariant.\\
The first set of invariances is associated with the solution of
the constraint $\Db_\ad\F=0$. It is given by
\bea
&\d\c=\Db^\ad\cb_\ad~~~,\non\\
&\d\sba_\ad=-\cb_\ad[m+\widetilde{P}(\F)]~~~.
\label{gaugechir}
\eea
The second set of invariances is
\bea
\d\c&=&0~,\non\\
\delta \sigma^\alpha&=& D_\b \sigma^{(\b\alpha)}~,\nonumber\\
\delta \sigma^{(\b\alpha)}  &=& D_\gamma \sigma^{(\gamma\b\alpha)}~,
\nonumber\\
\delta \sigma^{(\gamma\b\alpha)}  &=& D_\delta
\sigma^{(\delta\gamma\b\alpha)}~,
\nonumber\\
~&\vdots&~\non\\
\d\s^{(\a_n\a_{n-1}\cdots\a_1)}&=&D_{\a_{n+1}}\s^{(\a_{n+1}\a_n\a_{n-1}
\cdots\a_1)}~,\non\\
~&\vdots&
\label{gaugetr}
\eea  
The $\s^\a$ part of (\ref{gaugetr}) was studied in 
\cite{Q.F.T.S.NM,Sigma+PropLinear} where the quantization of the massless 
complex linear superfield was performed. 

We can gauge--fix these invariances in two steps.\\
First we consider the transformations (\ref{gaugechir}). We use the 
well known gauge--fixing procedure used for the massless scalar chiral 
superfield \cite{SUPERSPACE}. 
This amounts to adding a gauge--fixing term \cite{SUPERSPACE} which brings  
the operator $D^2\Db^2$ to $\Box$, and then, in our case, 
$(D^2\Db^2-|m|^2)$ to $(\Box-|m|^2)$ invertible also for $m=0$. 
Furthermore, the ghost fields introduced by this gauge--fixing completely
decouple from the physical fields \cite{SUPERSPACE}.\\
As a second step we consider the transformations (\ref{gaugetr}).
Since $\c$ does not transform, we can use the gauge--fixing procedure 
described in \cite{Q.F.T.S.NM,Sigma+PropLinear} for the case of a pure complex 
linear superfield. This procedure makes the operator in  
$\int d^8z~\s_\a D^\a\Db^\ad\sba_\ad$ invertible.
The gauge--fixing is developed by introducing an infinite tower of ghosts 
\cite{Q.F.T.S.NM} according to a superspace version of the 
Batalin--Vilkovinsky formalism. Furthermore, in \cite{Sigma+PropLinear} it was
proved that the tower of ghosts can be completely decoupled from the $\s_\a$ 
and $\sba_\ad$ fields by a finite number of ghost fields redefinitions. 
The same procedure can be applied without modifications to our case. The net 
result is the conversion of the operator $D^\a\Db^\ad$ into
the invertible operator $W^{\a\ad}$, the explicit expression of which was 
given in \cite{Sigma+PropLinear}.

What is important is that at the end of the two gauge--fixing 
procedures the ghosts introduced completely decouple from $\s_\a$, 
$\sba_\ad$, $\c$ and $\cb$, while the interaction terms are not modified. 
The gauge--fixed kinetic action is then
\beq
S^0_{CNM}+S^{tot}_{GF}=\int d^4x d^4\te~ (~\cb~,~\s_\a~)\(\begin{matrix}
\bbox-|m|^2&-\mb\Db^\ad\\-mD^\a&-W^{\a\ad}\end{matrix}\)
\(\begin{array}{c}\c\\ \sba_\ad\end{array}\)~.
\label{StotCNMm}
\eeq
Since the inverse of $W^{\a\ad}$ is \cite{Sigma+PropLinear}
\bea
W^{-1}_{\a\ad}&=&-{i\pa_{\a\ad}\over\Box}~+~
{3(kk'_1)^2+4-2{k'_1}^2\over 4(kk'_1)^2}i\pa_{\a\ad}{\Db^2D^2\over\Box^2}~+~
\non\\
&&~~~~~~~~~~~~~~~~~+~
{3k^2-2\over4k^2}i\pa_{\a\ad}{D_\b\Db^2D^\b\over\Box^2}~+~
{2-k^2\over 4k^2}i\pa_{\a\bd}i\pa_{\b\ad}{\Db^\bd D^\b\over\Box^2}~,
\eea
where $k$ and $k_1'$ are two gauge--fixing parameters, 
it is possible to invert
the kinetic operator in (\ref{StotCNMm}) and find the following propagators
\beq
\(\begin{matrix}
~~<\c\cb>&<\c\s_\a>\\&&\\~~<\sba_\ad\cb>&<\sba_\ad\s_\a>
\end{matrix}\)
=\(\begin{matrix}
~~-\frac{1}{\Box}\(1+{|m|^2\over\Box-|m|^2}{D^2\Db^2\over\Box}\)&&
{\mb\over\Box}\({D_\a\Db^2\over\Box}-\frac{1}{2}{\Db^2D_\a\over\Box}\)
\\&&&\\~~ -{m\over\Box}\({D^2\Db_\ad\over\Box}-
\frac{1}{2}{\Db_\ad D^2\over\Box}\)&&
W^{-1}_{\a\ad}-{|m|^2\over\Box}W^{-1}_{\b\ad}D^\b\Db^\bd W^{-1}_{\a\bd}
\end{matrix}\)~.
\label{propR}
\eeq
We observe that, in the limit $m=0$, the resulting $<\c\cb>$ and 
$<\sba_\ad\s_\a>$ propagators are exactly those known for the massless 
chiral and complex linear superfields, as expected.

\section{One--loop effective potential for CNM SYM theories}

We now look at the $N=1$ Super--Yang--Mills model described by the classical 
action
\beq
S=\int d^8z\[\Fib e^V\F-\Sb e^V\S\] +\frac{1}{4}\int d^6z~
\Tr\ W^\a W_\a +\frac{1}{4}\int d^6\bar{z}~
\Tr\ \bar{W}^\ad \bar{W}_\ad~~~,
\label{SCNMYM1}
\eeq
which is the CNM generalization of the SYM model with massless complex 
linear matter superfields studied in \cite{NM.S.Y.M.}.
The CNM matter superfields $\F^i$ and $\S^i$ satisfy $\Db_\ad\F=0$ and 
$\Db^2\S=m\F$, where $m$ is a gauge singlet. Thus both $\F$ and $\S$ belong
to the same representation of a gauge group. The vector superfield is in the
adjoint representation $(V)_{ij}\equiv V^a (T_a)_{ij}$ with $(T_a)_{ij}$ 
the Lie algebra generators in the representation of $\F$ and $\S$. 

Our aim is to compute the one--loop K\"ahler effective potential for the matter
superfields. 
The one--loop divergent terms come from the contributions 
which have external $\F$, $\Fib$, $\S$ and $\Sb$ fields without any spinorial 
and space--time derivatives acting on them. We focus on this kind of diagram.
  
In order to proceed we perform the quantum--background splitting 
$\S\to\S_Q+\S_B$, $\F\to\F_Q+\F_B$ and require that $\Db^2\S_Q=m\F_Q$ and 
$\Db^2\S_B=m\F_B$, even if the latter condition is not strictly necessary for 
the computation we are going to perform.\\ 
Inserting the splitting into the action (\ref{SCNMYM1}), in addition 
to the ordinary kinetic terms for $\F$, $\S$ and the gauge fields,
 we find the one--loop relevant interaction terms
\bea
\int d^8z\[\(\Fib_B V\F_Q+\Fib_Q V\F_B+ \frac{1}{2}\Fib_B V^2\F_B\)+
\(-\Sb_B V\S_Q-\Sb_Q V\S_B-\frac{1}{2}\Sb_B V^2\S_B\)\]+\cdots~.~
\label{SCNMYM2}
\eea
Since we are considering $\Db^2\S_Q=m\F_Q$, using the quantization procedures 
of the previous section, it is possible to find the effective propagators for 
the physical superfields $\F_Q$, $\Fib_Q$, $\S_Q$ and $\Sb_Q$ 
\bea
<\F_Q^i\Fib_Q^j>~=~-\d^{ij}{\Db^2D^2\over\bbox-|m|^2}\d^8(z-z')&,&
<\S_Q^i\Fib_Q^j>~=~-\d^{ij}{mD^2\over\bbox-|m|^2}\d^8(z-z')~,\non\\\non\\
<\F_Q^i\Sb_Q^j>~=~-\d^{ij}{\mb\Db^2\over\bbox-|m|^2}\d^8(z-z')&,&
<\S_Q^i\Sb_Q^j>~=~\d^{ij}\[1-{D^2\Db^2\over\bbox-|m|^2}\]\d^8(z-z')
~.~~~\label{Matter}
\eea
Furthermore, we have the propagator for the vector superfield which, in the
Landau gauge, reads
\cite{SUPERSPACE,NM.S.Y.M.,kahlergaugechirale}
\beq
<V^aV^b>~=~\d^{ab}{D_\a\Db^2D^\a\over\bbox^2}\d^8(z-z')~~~.\label{vvG}
\eeq
Now, we calculate only one--loop amplitudes without derivatives on the 
external fields, using the above propagators.

It is convenient to consider the effective 
Yang--Mills propagators obtained by inserting the vertices 
$\frac{1}{2}\Fib_B V^2\F_B$ and $-\frac{1}{2}\Sb_B V^2\S_B$. 
Summing on $1$--PI diagrams we find
\bea
<<V^aV^b>>&=&\[\(\bbox-\widetilde{V}^{(4)}\)^{-1}\]^{ab}
{D_\a\Db^2D^\a\over\bbox}\d^8(z-z')~,
\label{vvGeff}
\eea
with
\beq
\widetilde{V}^{(4)}_{ab}\equiv\frac{1}{2}\[\(T_aT_b+T_bT_a\)_{ij}
(\Fib_B^i\F_B^j-\Sb_B^i\S_B^j)\]
=\frac{1}{2}\Big[[\Fib_B\F_B]_{ab}-[\Sb_B\S_B]_{ab}\Big]~~~,
\eeq
having defined $[AB]_{ab}\equiv\(T_aT_b+T_bT_a\)_{ij}(A^iB^j)$. The one--loop 
amplitudes can now be calculated considering the cubic vertices in 
(\ref{SCNMYM2}) and using $<<V^aV^b>>$ as the vector propagator.\\
For the calculation we are performing the matter propagators with at least one 
superfield $\F_Q$ or $\Fib_Q$ in (\ref{Matter}) are orthogonal to the 
YM--propagator (\ref{vvGeff}) and then terms built with 
the cubic vertices $\Fib_B^i (T_a)_{ij}V^a\F_Q^j+\Fib_Q^i(T_a)_{ij}V^a\F_B^j$ 
in (\ref{SCNMYM2}) are zero. Consequently, only diagrams constructed using the 
vertices $V^{(3)}=\Sb_B^i (T_a)_{ij}V^a\S_Q^j+\Sb_Q^i(T_a)_{ij}V^a\S_B^jV^a$
connected by the propagators $<<V^aV^b>>$, $<\S_Q^i\Sb_Q^j>$ contribute. 
After Fourier transforming, and summing up all the diagrams, we obtain
\beq
\Gamma^{(3)}=-\int {d^4p\over (2\pi)^4}d^4\te{2\over p^2}\Tr\ggl
\ln{\[1+{[\Fib_B\F_B]_{ab}+[\Sb_B\S_B]_{ab}\over 2p^2}\]}
-\ln{\[1+{[\Fib_B\F_B]_{ab}-[\Sb_B\S_B]_{ab}\over 2p^2}\]}
\ggr~~~.
\eeq
Beyond this contribution, we find another term obtained with only the vertices 
$\frac{1}{2}\Fib_B^i(T_aT_b)_{ij}V^aV^b\F_B^j$ and 
$-\frac{1}{2}\Sb_B^i(T_aT_b)_{ij}V^aV^b\S_B^j$ considered
\bea
\Gamma^{(4)}&=&-\int {d^4p\over (2\pi)^4}d^4\te{2\over p^2}
\Tr\ggl\ln{\[1+{[\Fib_B\F_B]_{ab}-[\Sb_B\S_B]_{ab}\over 2p^2}\]}
\ggr~~~.
\eea
Summing $\G^{(3)}$ and $\G^{(4)}$ we find the following divergent contribution 
to the effective potential
\bea
%\Gamma_{eff}&=&
-\int {d^4p\over (2\pi)^4}d^4\te{2\over p^2}\Tr\ggl
\ln{\[1+{[\Fib_B\F_B]_{ab}+[\Sb_B\S_B]_{ab}\over 2p^2}\]}\ggr~~~.
\label{Geffpre}
\eea
To cancel this divergence we introduce a counterterm which 
renormalizes the original matter fields action. 
Evaluating the momentum integrals, having introduced a renormalization mass 
$\mu$, we find for the renormalized effective potential
\beq
\Gamma_{eff}=\frac{-1}{(4\pi)^2}\Tr\[\([\Fib_B\F_B]+[\Sb_B\S_B]\)
\ln{\({[\Fib_B\F_B]+[\Sb_B\S_B]\over 2\mu^2}\)}\]~.~~~~~~
\label{Geffref}
\eeq

Some comments are now in order. We note that although we are working with a 
massive theory, our result does not depend on mass $m$. This is due to the 
fact that we have focused only on amplitudes without space--time and 
spinor derivatives acting on the external fields.
Furthermore, the mass term is not explicit in the action as it is defined using
the kinematic constraints.

The next, convergent, contributions to the one--loop K\"ahler effective 
action will result from terms having derivatives also on the external fields. 
For this kind of terms all the massive propagators would be relevant,
and since $\Db^2\S_B=m\F_B$ (and in general $\Db^2\S_B=Q(\F_B)$), these 
contributions depend explicitly on $m$ (and in general on the parameters 
of the potential $Q$).\\
It would be interesting to further pursue this investigation by considering
the nontrivial interactions  $\widetilde{P}(\F)$, $W(\F)$, $K(\F,\Fib,\S,\Sb)$,
studying in detail the structure of the classical vacua
around which the saddle point approximation should be developed, and extending 
the analysis to more than one--loop.

\section{Covariant quantization of CNM superfields and the Konishi anomaly}
\label{subsection:Q.C.CNM}

In section \ref{section:QCNM1} we developed a superspace formalism for 
quantizing the CNM multiplet with a generic potential $Q(\F)$. Now we study 
the coupling with background gauge fields. 
This can be done by introducing a completely covariant formalism with respect 
to both SUSY and gauge transformations. Given the set of covariant derivatives 
$\Dc_A=(\Dc_\a,\Dcb_\ad,\Dc_{\a\ad})$ \cite{SUPERSPACE,SBFM}, we define 
covariant chiral and covariant complex linear superfields as
\bea
\Dcb_\ad\F=0&,&\Dcb^2\S=Q(\F)~.
\label{CNMcovkin}
\eea
with $Q(\F)$ as in (\ref{QCNM}). 

We study the particular case $Q(\F)\equiv m\F$, $m\ne 0$.
We proceed here by choosing slightly different solutions of the 
kinematic constraints compared with 
(\ref{risvincoloCNM1}, \ref{risvincoloCNM2})
\bea
\F=\Dcb^2\c~~~~~~&,&~~~~~~\S=m\c~~~,
\label{covrisvinCNM1}\\
\Fib=\Dc^2\cb~~~~~~&,&~~~~~~\Sb=\mb\cb~~~\label{covrisvinCNM2}.
\eea
This is a natural consequence of the fact that, apart from the 
constant $m$, the present constraint $\Dcb^2\S=m\F$ simply identifies $\S$ with
the generic superfield $\c$ which solves the chiral constraint 
$\Dcb_\ad\F=0$ \cite{BuchbinderKuzenko}.
In other words, since $\widetilde{P}(\F)\equiv0$ and $m\ne0$, 
we can absorb the $\sba_\ad$ ($\s_\a$) superfield in (\ref{risvincoloCNM1},
\ref{risvincoloCNM2}) by redefining the $\c$ ($\cb$) superfield as 
$\c +\frac{1}{m}\Dcb^\ad\sba_\ad$ ($\cb +\frac{1}{\mb}\Dc^\a\s_\a$) obtaining
(\ref{covrisvinCNM1}, \ref{covrisvinCNM2}).

Using (\ref{covrisvinCNM1}, \ref{covrisvinCNM2}), the 
action $\int d^8z\[\Fib\F-\Sb\S\]$ 
gives a kinetic quadratic term
\beq
S_{kin(m\ne0)}=\int d^8z~\cb\(\Dc^2\Dcb^2-|m|^2\)\c~,
\eeq
which has an invertible kinetic operator; no gauge--fixing procedure is now 
necessary. The $<\c\cb>$ covariant propagator is then
\beq
<\c\cb>={1\over|m|^2}\[1-\Dc^2{1\over\Box_+-|m|^2}\Dcb^2\]~,
\label{ccbProp}
\eeq
where
\bea
\Box_+\equiv\Dc^2\Dcb^2+\Dcb^2\Dc^2+\Dcb_\ad\Dc^2\Dcb^\ad=
\frac{1}{2}\Dc^{\a\ad}\Dc_{\a\ad}-iW^\a\Dc_\a-i\frac{1}{2}(\Dc^\a W_\a)
\eea
and $W_\a\equiv i\frac{1}{2}[\Dcb^\ad,\{\Dc_\a,\Dcb_\ad\}]$ is the gauge field 
strength \cite{SUPERSPACE,SBFM}.

We note that (\ref{ccbProp}) is not defined when $m\equiv0$. This 
is a natural consequence of the fact that, as previously said, solutions 
(\ref{covrisvinCNM1}, \ref{covrisvinCNM2}) are suitable for treatment of the 
kinetic term associated with the CNM multiplets only when $m\ne0$ and 
$Q(\F)\equiv m\F$.
In the case of generic $Q(\F)$ a covariant generalization of solutions 
(\ref{risvincoloCNM1}, \ref{risvincoloCNM2}) should be used to develop the
quantization. 
This would require the construction of an explicitly covariant generalization
of the quantization procedure of section \ref{section:QCNM1}. In particular,
a suitable procedure would have to be developed to treat the ghosts
in a manifestly covariant way.

It is possible to explicitly study the quantization of the present CNM 
model using the physical covariant superfields $\F$, $\Fib$, $\S$ and $\Sb$. 
In fact, the effective propagators are 
\bea
<\Fib\F>=\Dc^2<\cb\c>\Dcb^2=-\Dc^2{1\over\Box_+-|m|^2}\Dcb^2&,&
<\S\Fib>=m<\c\cb>\Dc^2=-{m\Dc^2\over\Box_--|m|^2}~~~,
\non\\
\non\\
<\S\Sb>=m<\c\cb>\mb=1-\Dc^2{1\over\Box_+-|m|^2}\Dcb^2
&,&
<\F\Sb>=\Dcb^2<\c\cb>\mb=-{\mb\Dcb^2\over\Box_+-|m|^2}~~~.\non\\
\label{covpropCNM}
\eea
We observe from (\ref{covpropCNM}) that, in this case, the physical 
propagators are also well defined when $m\equiv0$. This leads to the 
conjecture that the covariant $<\S\Sb>$ propagator for the massless complex 
linear superfield should be $<\S\Sb>=1-\Dc^2{1\over\Box_+}\Dcb^2$. 

We now apply the covariant quantization formalism to study the
Konishi anomaly in CNM SYM theories. We consider a pair of covariantly 
CNM superfields $\F$ and $\S$ satisfying (\ref{CNMcovkin}) with 
$Q(\F)\equiv m\F$.
The kinetic action for the CNM theory has the form of (\ref{SCNM}). 
Beyond the gauge invariance, action (\ref{SCNM}), together with the
constraint $\Dcb^2\S=m\F$, has the global invariance 
\beq
(\F,\S)\to\exp{\[i\a\]}(\F,\S)~~~,~~~(\Fib,\Sb)\to\exp{\[-i\a\]}(\Fib,\Sb)~~~.
\label{ftrasf}
\eeq
It is important to note that, due to the kinematic constraints,
we are forced to choose the same charges for $\F$ and $\S$.
The resulting current $J_0=(\Fib\F-\Sb\S)$ satisfies the classical 
conservation equation 
\beq
\Db^2\(\Fib\F-\Sb\S\)=0~~~.
\label{ClassicK}
\eeq
We observe that, in the case of SYM theories with pure 
chiral matter multiplets, the $U(1)$ symmetry $\F\to\exp{\[i\a\]}\F$ 
presents the chiral anomaly known as the Konishi anomaly \cite{Konishi}.
In \cite{CNMmassless} it was observed that the phase trasformation 
(\ref{ftrasf}) acts on the Dirac spinor (\ref{diracCNM}) as a pure vector 
transformation $\Psi_{CNM}\to\exp{[i\a]}\Psi_{CNM}$, and so it was argued that 
the CNM theory should be anomaly free.
  
Now, as a simple application of the covariant quantization formalism previously
developed, we derive this property explicitly.
As in the pure chiral case \cite{Konishi}, we are led to study potential
anomalies for the composite invariant gauge operators $\Sb\S$ 
and $\Fib\F$. We separately study the expectation values of the operators
$\Db^2(\Sb\S)$ and $\Db^2(\Fib\F)$. In particular, we expect
\bea
\left<\Db^2(\Sb\S)(z)\right>=m\left<(\Sb\F)(z)\right>+{\rm NM.Anomaly}~,~
\left<\Db^2(\Fib\F)(z)\right>=m\left<(\Sb\F)(z)\right>+{\rm C.Anomaly}~,~
\eea
where the anomaly terms are quantum corrections to the classical equations for 
the composite operators.
We follow \cite{Konishi, SUPERSPACE} and regularize the composite operators 
$\Sb\S$ and $\Fib\F$, using the Pauli--Villars regularization. 
In particular, as for the pure chiral case, it is possible to regularize the 
theory by introducing pairs of CNM covariant superfields connected by the 
kinematic constraints $\Dcb_\ad\F_M=0$ and $\Dcb^2\S_M=M\F_M$, with the 
parameter $M$ satisfying $M\gg m$. At this point, we consider the regularized 
amplitudes
\bea
<\Fib\F(z)>-<\Fib_M\F_M(z)>~~~~~~{\rm and}~~~~~~<\Sb\S(z)>-<\Sb_M\S_M(z)>~~~,
\eea
and compute
\beq
\lim_{M\to\infty}\Dcb^2(<\Sb\S(z)>-<\Sb_{M}\S_{M}(z)>)~~~{\rm and}~~~
\lim_{M\to\infty}\Dcb^2(<\Fib\F(z)>-<\Fib_{M}\F_{M}(z)>)~~~.
\eeq
Using the effective massive covariant propagators (\ref{covpropCNM}), we find 
that these terms are both equal to
\bea
\lim_{M\to\infty}
\Tr\int d^8z'~\d^8(z'-z)\ggl\Dcb^2\Dc^2\frac{-1}{\Box_+ -|m|^2}
\Dcb^2\d^8(z-z')
+\Dcb^2\Dc^2\frac{1}{\Box_+ -M^2}\Dcb^2\d^8(z-z')\ggr=\non\\
=\lim_{M\to\infty}
\Tr\int d^8z'~\d^8(z'-z)\ggl\frac{-|m|^2}{\Box_+ -|m|^2}\Dcb^2\d^8(z-z')+
\frac{M^2}{\Box_+ -M^2}\Dcb^2\d^8(z-z')\ggr=\non\\
=m<\F\Sb(z)>+\lim_{M\to\infty}\Tr\int d^8z'~\d^8(z'-z)\ggl
\frac{M^2}{\Box_+ -M^2}\Dcb^2\d^8(z-z')\ggr~.
\eea
Then, the anomaly term is
\beq
\lim_{M\to\infty}\ggl M^2\Tr\int d^8z'~\d^8(z'-z)\frac{\Dcb^2}{\Box_+ -M^2}
\d^8(z-z')\ggr~~~.
\label{Anomaly}
\eeq
Once the $\Dc$--algebra and the integral have been performed, 
we find that the anomaly term (\ref{Anomaly}) 
is exactly $-\frac{1}{32\pi^2}\Tr\[W^\a W_\a\]$, the same as the Konishi 
anomaly\footnote{For references on the covariant formalism in chiral theories 
and an explicit calculation of (\ref{Anomaly}) see \cite{SBFM} and 
\cite{SUPERSPACE,SBFMC}.}.\\ 
The anomalous equations for the gauge invariant composite operators $\S\Sb(z)$
and $\Fib\F(z)$ are then
\bea
\Db^2\left<\Sb\S(z)\right>&=&m\left<\Sb\F(z)\right>
-\frac{1}{32\pi^2}\Tr\[W^\a W_\a\]~~~,\\
\label{anomalyCNM3}
\Db^2\left<\Fib\F(z)\right>&=&m\left<\Sb\F(z)\right>
-\frac{1}{32\pi^2}\Tr\[W^\a W_\a\]~~~.
\label{anomalyCNM4}
\eea
Therefore, there is a complete cancellation between the chiral
and nonminimal anomalies in the conservation equation for the current $J_0$
(\ref{ClassicK}) which is found to be anomaly free, as expected.

\section{Glueball superpotential in N=1 CNM SYM}

Recently, some new insights have been obtained into the non--perturbative 
dynamics of $N=1$ supersymmetric gauge theories constructed by chiral matter 
superfields.
In particular Dijkgraaf and Vafa have argued the existence of a connection 
between the 't Hooft limit sector of zero dimensional bosonic matrix models 
and the effective Wilsonian holomorphic superpotential 
for a wide class of $N=1$ supersymmetric gauge theories. The relation has been
originally conjectured by using a chain of dualities in string theories 
\cite{DV}. Subsequently 
it has been proved by using covariant supergraph 
tecniques \cite{DGLVZ}. In \cite{CDSW}, the connection has been discussed by
using generalized Konishi anomaly equations having the same structure of the 
loop equations of a matrix model \cite{DV,CDSW}.
Both these approaches determine the superpotential up to a term given by 
the Veneziano--Yiankielowicz superpotential which takes care of pure gauge 
dynamics. 

Inspired by these interesting results it is natural to ask if and how $N=1$ 
supersymmetric gauge theories defined by CNM matter multiplets present 
low--energy dynamical properties similar to the pure chiral case studied on 
the grounds of the Dijkgraaf--Vafa conjecture.\\
In this section we discuss some simple properties of the glueball 
superpotential in $N=1$ CNM SYM theories. In particular, we concentrate on the
simplest class of CNM multiplet defined by $\Dcb^2\S=Q(\F)\equiv m \F$, 
$m\ne 0$. Furtheremore, we consider the following interaction term
\beq
\int d^4xd^2\te\ W(\F)+{\rm h.c.}~~~,
\label{chiralInteractions}
\eeq
where $W(\F)$ is a gauge invariant superpotential at least cubic in the chiral
superfields $\F$. For these class of theories, in the previous section, we 
have constructed the covariant quantization. We discuss a perturbative 
approach to the calculation of the glueball superpotential by using covariant 
supergraph techniques as in \cite{DGLVZ}. We want to integrate out the massive
degrees of freedom and find the effective superpotential for the gauge 
superfields. We consider the simplest case with unbroken gauge group. Then, the
perturbative part of the effective superpotential $\int d^2\te W_{pert}$ is 
obtained from vacuum amplitudes where only CNM matter fields propagate 
considering the gauge superfields as a background.
Using the analysis of \cite{DGLVZ,CDSW}, we can argue that the perturbative
part of the superpotential have the form
\beq
\int d^2\te~W_{pert}[S,w_\a,(\cdots)]~~~~~~{\rm with}~~~~~~
S={1\over 32\pi^2}\Tr~W^\a W_\a~~~,~~~w_\a={1\over 4\pi}\Tr~W_\a~~~,
\label{Weff}
\eeq
where ($\cdots$) indicate the dependence of $W_{pert}$ from the matter 
coupling constants. The form of (\ref{Weff}) 
follows from the fact that $W_{eff}$, by definition, is an element of the 
chiral ring \cite{CDSW} which in this case is generated by $S$ and $w_\a$. 
We remember that for pure chiral theories it is well known that the
superpotential is holomorphic in the matter coupling constants
\cite{Seiberg9309335,DGLVZ,CDSW}. At the moment we have not imposed any 
restriction, as holomorphicity, in the CNM case. We now discuss this point.

The covariant propagators for the CNM matter superfields are give in 
(\ref{covpropCNM}). Considering that the interactions 
(\ref{chiralInteractions}) are defined by the superfields $\F$ and $\Fib$ 
alone, it is clear that all the perturbative contributions involve only 
the $<\Fib\F>$ covariant propagator in (\ref{covpropCNM}).  
Therefore, from the perturbative point of view the considered CNM theories has
the same structure of pure massless chiral theories with superpotential 
$W(\F)$ \cite{SUPERSPACE}. Using this observation we can deduce that
holomorphicity perturbatively works also in these CNM models being a property 
of the pure chiral theories. Furthermore, since there are no holomorphic 
propagators $<\F\F>$ in our case and all the perturbative contribution will be
necessary non--holomorphic, it is then simple to observe that there are no 
perturbative contributions to the glueball superpotential in these CNM 
theories.\\
It is interesting to note that we have not explicited the representation in 
which the CNM matter multiplets are. Anyway, the above arguments, being 
associated to the covariant perturbative structure of CNM models, should not 
depend on the matter fields representation.\\ 
Therefore, it seems that from the point of view of the low--energy gauge 
dynamics these CNM theories look very different from the pure chiral theories 
studied on the grounds of the Dijkgraaf--Vafa conjecture \cite{DV,DGLVZ,CDSW}. 

We observe that in order to derive the above result we have assumed unbroken 
gauge group and $Q(\F)\equiv m\F$. It would be very interesting to extend the 
analysis of this section to the general case. To this respect it should be 
found a generalization of the covariant quantization of section $5$.

\section{Conclusions and further issues}

In this letter we have studied some quantum properties of $N=1$ supersymmetric
field theories with chiral/nonminimal scalar multiplets defined by
a chiral ($\F$) superfield and a complex linear ($\S$) superfield 
kinematically coupled by $\Db^2\S=Q(\F)$.

In particular, generalizing the quantization techniques typical of massless
chiral and complex linear superfields, we have developed the superspace 
quantization for CNM models with a general potential $Q(\F)$.  
When $Q(\F)\equiv m\F$ and the CNM scalar multiplets are coupled to background
gauge fields, we have also constructed the covariant quantization formalism.

As preliminary applications, in sections $4$, $5$ and $6$ we have studied some 
quantum properties of supersymmetric gauge theories built using CNM 
scalar matter superfields. In the particular case $Q(\F)=m\F$ we have computed
the one--loop contribution to the effective action for the matter superfields,
proved the anomaly free nature of these models, and discussed some properties
of the glueball superpotential.

There are several classical and quantum issues in the study of 
CNM models that we hope to clarify in the future.
In particular, the dynamical properties of CNM theories with nontrivial 
kinematic interaction $\widetilde{P}(\F)$ (\ref{QCNM}) call for further 
investigation. 
For example, we would like to clarify the precise structure of the chiral 
sector of CNM theories and see exactly how the fact that on--shell $\Sb$ is a 
chiral superfield influences the quantum properties of these theories.
Looking in this direction we should understand whether the amplitudes in the 
chiral ring ($F$--terms) \cite{DGLVZ,CDSW} are holomorphic in the tree--level 
coupling constants of the chiral potentials $Q(\F)$ (\ref{QCNM}) and 
$W(\F)$ (\ref{Sint}). Holomorphicity seems to be valid in the CNM case as in 
the pure chiral, since the potentials $Q(\F)$ and $W(\F)$ are only defined in 
terms of chiral superfields and it is then possible to argue holomorphicity 
using arguments of naturalness widely used in pure chiral theories 
\cite{Seiberg9309335}.\\
Furthermore, we would like to find a generalization of the covariant 
quantization formalism of section $5$ for CNM multiplet defined with generic 
potential $Q(\F)$. Then, it will be interesting to extend the analysis of 
sections $5$ and $6$.

Finally, once the properties of CNM models constructed giving mass only 
kinematically are clearer, it will be interesting to study models
with both CNM and chiral mass terms. The quantization procedure of this letter
can be easily generalized to this case \cite{tesiGTM}.

\section*{Acknowledgements}
\noindent The author would like to thank Silvia Penati for suggesting the 
problem and for helpful discussions and suggestions.

\end{document}